**Propagation control of octahedral tilt in SrRuO$_3$ via artificial heterostructuring**

*Seung Gyo Jeong, Gyeongtak Han, Sehwan Song, Taewon Min, Ahmed Yousef Mohamed, Sungkyun Park, Jaekwang Lee, Hu Young Jeong, Young-Min Kim, Deok-Yong Cho, and Woo Seok Choi\**

S. G. Jeong, Prof. W. S. Choi
Department of Physics, Sungkyunkwan University, Suwon 16419, Republic of Korea
E-mail: choiws@skku.edu

G. Han, Prof. Y.-M. Kim
Department of Energy Sciences, Sungkyunkwan University, Suwon 16419, Korea

G. Han, Prof. Y.-M. Kim
Center for Integrated Nanostructure Physics, Institute for Basic Science, Suwon 16419, Korea

S. Song, T. Min, Prof. S. Park, Prof. J. Lee
Department of Physics, Pusan National University, Busan 46241, Korea

A. Y. Mohamed, Prof. D.-Y. Cho
IPIT & Department of Physics, Jeonbuk National University, Jeonju 54896, Republic of Korea

Prof. H. Y. Jeong
UNIST Central Research Facilities and School of Materials Science and Engineering, Ulsan National Institute of Science and Technology, Ulsan 44919, Korea

Bonding geometry engineering of metal-oxygen octahedra is a facile way of tailoring various functional properties of transition metal oxides. Several approaches, including epitaxial strain, thickness, and stoichiometry control, have been proposed to efficiently tune the rotation and tilting of the octahedra, but these approaches are inevitably accompanied by unnecessary structural modifications such as changes in thin-film lattice parameters. In this study, we propose a method to selectively engineer the octahedral bonding geometries, while maintaining other parameters that might implicitly influence the functional properties. A concept of octahedral tilt propagation engineering has been developed using atomically designed SrRuO$_3$/SrTiO$_3$ superlattices. In particular, the propagation of RuO$_6$ octahedral tilting within the SrRuO$_3$ layers having identical thicknesses was systematically controlled by varying the thickness of adjacent SrTiO$_3$ layers. This led to a substantial modification in the electromagnetic properties of the SrRuO$_3$ layer, significantly enhancing the magnetic moment of Ru. Our approach provides a method to selectively manipulate the bonding geometry of strongly correlated oxides, thereby enabling a better understanding and greater controllability of their functional properties.

Recent developments in atomic-scale precision epitaxy and microscopy of transition metal oxides have rediscovered the importance of local atomic coordination in the determination of their physical properties.[1-7] In $A$BO$_3$ perovskites, besides the conventional lattice degree of freedom, such as lattice parameters, octahedral distortions (tilt and rotation), are also being considered to be an accessible degree of freedom in the context of modifying the opto-electronic and magnetic properties of such materials. Indeed, the transition metal-oxygen ($M$-O) bonding geometry is closely coupled to the corresponding charge, spin, and orbital states, leading to adjustable functionalities of oxides. In particular, the directional hybridization of localized $d$-electrons in $B$-site transition metals with oxygen $p$-orbitals modifies their crystalline symmetries, which further breaks the degeneracy. For example, in La$_{1-x}$Sr$_x$MnO$_3$ thin films, electromagnetic phase transitions were induced by $x$-dependent modifications in the octahedral network.[8] In $A$NiO$_3$, a metal-insulator transition, coupled with a magnetic transition, was achieved by decreasing the ionic radii of the $A$-sites, which resulted in a decrease in the Ni-O-Ni bond angle from 180°.[9, 10]

SrRuO$_3$ (SRO) is a prototypical material used to study the $M$-O bonding geometry tuning of functional properties. Emergent phenomena such as metal-insulator transitions, superconductivity, strong magneto-structural couplings, tunable topological phases, and enhanced electrocatalytic activities have been reported to be strongly dependent on the nature of the Ru-O bond.[11-17] In bulk, SRO is orthorhombic with the $Pbnm$ space group.[18] It is an itinerant ferromagnet (FM) with a nearly half-metallic property, whose electronic state can be precisely determined by customizing its octahedral distortion.[19] For instance, the tunable electromagnetic ground state can be manipulated to enhance the spin-polarized current for spintronics.[20]

The engineering of $RuO_6$ octahedral distortion in SRO has been achieved via various approaches, including epitaxial strain modification using different substrates, thickness control, addition of buffer/capping layer, and stoichiometry control (Figure 1). According to a computational study, both orthorhombic (with octahedral tilt) and tetragonal (without octahedral tilt) structures of SRO are nearly degenerate in energy (with a difference of only a few tens of meV), leading to a facile control over their octahedral tilts (Figure 1a).[21] Conventionally, the lattice mismatch between the thin film and the substrate imposes an epitaxial strain. Yet, the octahedral distortion of the substrate can impose an additional geometric constraint, based on the continuity of *M*-O-*M* bonds across the hetero-interface. Figure 1b depicts the substrate dependence of SRO thin films with a controllable $RuO_6$ octahedral distortion. On a $GdScO_3$ substrate, which possesses an octahedral distortion analogous to that of bulk SRO, the SRO layer naturally maintains its original octahedral distortion.[22] In contrast, the cubic symmetry of $SrTiO_3$ (STO) suppresses the distortion.[19, 23] Further, as substrate-induced modification (or interfacial coupling) of $RuO_6$ distortion cannot prevail over tens of unit cells,[24] thickness-dependent transition has been accomplished (Figure 1c).[23, 25] More recently, octahedral tilt engineering was executed by inserting an additional buffer (capping) layer below (above) the thin film (Figure 1d).[26-28] Finally, stoichiometry (Sr and O vacancies) engineering can also alter the octahedral distortion and crystalline symmetry of SRO thin films.[29] While these approaches have been successful in modifying the crystalline symmetry of SRO, unintended effects originating from dissimilar substrates, partial strain relaxation, electronic charge transfer at the interface, and thickness-

(or composition-) dependent modifications of the electronic structures, could obscure the intrinsic understanding of the role of engineered *M*-O bond geometry.

In this paper, we report controllable octahedral tilt propagation by atomically designing artificial superlattices (SLs). Conventional octahedral modifications of SLs were achieved by changing the active layer.[30, 31] However, we were able to achieve the octahedral modification of the active layer within the SL by changing the inactive layer.[32] We maintained the same substrate (STO) to keep the degree of epitaxial strain constant, as well as the identical thickness and stoichiometry of the SRO layers. Yet, it is possible to selectively control the octahedral bonding geometry. The SRO/STO SL system was chosen as the *A*-site ion (Sr) was not disturbed and charge transfer across the hetero-interface was effectively suppressed.[33, 34] While a recent publication shows magnetic anisotropy change in SRO/STO SLs,[35] a structural phase transition has not been reported so far, depending on the STO thickness. Figure 1f shows the key features of our approach, including the dependence of $RuO_6$ octahedral tilt on the thickness of the STO layer within the SL. The cubic nature of the STO layer was observed to restrain the octahedral tilt of the SRO layer. Hence, the thin STO layer allowed more efficient propagation of the octahedral tilt than the thick STO layer. It was concluded that the atomic-scale periodicity of the SL governs structural propagation across the entire SL and eventually determines the macroscopic crystalline symmetry and resultant electromagnetic ground state.

SRO/STO SLs with modulated octahedral distortions were realized using the atomic-scale precision growth of pulsed laser epitaxy (PLE). Figure 2 shows the precisely controlled atomic unit cell (u.c.) layers of SRO and STO, especially for the $[(SRO)_{\alpha=6}|(STO)_{\beta}]_{10}$ ($\alpha = 6$ u.c. layers of SRO and $\beta$ u.c. layers of STO repeated 10 times along the growth direction, $[6|\beta]$) SL series on a single-crystalline (001) STO substrate. X-ray diffraction (XRD) $\theta$-$2\theta$ scans (Figure 2a) and reciprocal space maps (Figures 2b and S2) showed coherent SL peaks corresponding to the periodicity of each sample, which was fully strained to the substrate. The in-plane strain could be maintained owing to repeated clamping of the SRO layer by the STO layer. The crystalline structure of orthorhombic SRO has been schematically shown in the inset of Figure 2c, in which orthorhombicity is defined to be $a_o/b_o$ (Here, the subscript "o" represents the orthorhombic lattice). Orthorhombic distortion also leads to the tilting of the *M*-O-*M* bond angle ($\theta_{M\text{-}O\text{-}M} = 167°$ for bulk SRO).[11, 36] The ratio, $a_o/b_o$ was macroscopically characterized using off-axis XRD $\theta$-$2\theta$ scans around the (204) STO plane (Figure S1),[37] and the results have been summarized in a structural phase map, as functions of the thicknesses of the STO and SRO layers, in Figure 2c. Corresponding to SRO layers with a thickness ($\alpha$) less than ~4 u.c., the SLs did not exhibit any octahedral distortion; hence, the tetragonal SRO phase was consistently stabilized, irrespective of the thickness of the STO layer ($\beta$). This result is consistent with recent SRO/STO SLs study. On the other hand, when $\alpha \geq 8$ u.c., orthorhombic symmetry was maintained irrespective of $\beta$, although $a_o/b_o$ was observed to systematically decrease with an increase in $\beta$ (at least up to $\beta = 8$ u.c.). When $\alpha = 6$ u.c., a surprising $\beta$-dependent structural phase transition was detected in the SLs, i.e., the phase was observed to be orthorhombic when $\beta \leq 4$ u.c. but tetragonal when $\beta \geq 6$ u.c. Furthermore, the coherent tetragonal SRO could be stabilized up to ~120 nm of SRO thickness with $\beta = 8$ u.c. (Figure S3), providing another advantage of our strategy of propagation control of octahedral tilt.

The unprecedented $\beta$-dependency of octahedral tilt penetration was microscopically visualized using scanning transmission electron microscopy (STEM). Figures 2d,e show the high-angle annular dark-field (HAADF) (left) and annular bright-field (ABF) (right) STEM images of [6|$\beta$] SLs with $\beta = 2$ and 8, respectively. The images correspond to the cross-sectional pseudocubic (100) plane. The HAADF-STEM images showed the coherent atomic arrangements in the SRO/STO SLs. It is to be noted that each interface had a thickness deviation of less than 1 u.c. (~0.4 nm), which might have originated from the step-and-terrace structure of the substrate and the thin film. Note that < 1 u.c. deviation of the SRO layer thickness for sufficiently large thickness ($\alpha > 3$) does not influence its electromagnetic properties significantly. The atomic positions of the oxygen ions were clearly detectable with sub-Å precision as dark features in the ABF-STEM images. Hence, quantitative octahedral distortions along the out-of-plane direction were extracted (also see Figure S4). As exemplified in Figure 2f, the oxygen octahedral distortions prevailed within the SRO layers for the SL with $\beta = 2$. $\theta_{M\text{-}O\text{-}M}$ was measured to be minimal at ~175° at the center of the SRO layer and to gradually increase to ~179° towards the interface with STO. Meanwhile, the SL with $\beta = 8$ exhibited a highly suppressed distortion of ~1° (Figure 2f). The stark discrepancy between the two cases can be attributed to the competition between the cubic symmetry of the STO layer and the octahedral distortion of the SRO layer within the SL.

The suppressed octahedral tilt in the SRO layer led to enhanced magnetic exchange interactions between the Ru ions. As shown in Figure 3, we characterized the magnetic properties of the SLs along the out-of-plane direction, which corresponds closely to the magnetic easy-axis of typical SRO thin films.[38] Field-cooled temperature-dependent magnetization ($M(T)$) revealed characteristic FM behavior with a critical transition temperature of $T_c$ = ~140 K (Figure 3a). The $T_c$ values of the SLs were measured to be lower than those of the single SRO thin films (~30 nm), owing to the diminished FM interaction in the atomically thin SRO layers (e.g., 6 u.c.). These results were consistent with previous theoretical and experimental observations.[34, 39-41] Even among the SLs of identical SRO thickness, however, the tetragonal phases ($\beta \geq 6$ u.c.) exhibited higher $T_c$ values compared to that of the orthorhombic phases ($\beta \leq 4$ u.c.) (Figure 3c). The result pinpoints that the tetragonal symmetry is favorable in the enhancement of ferromagnetic exchange, consistent with a recent prediction based on density functional theory (DFT) calculation.[28] Further, we also noted a systematic increase in $T_c$ with an increase in $β$ within the SLs with tetragonal SRO layer, of which the exact origin is unclear. A possible scenario can be implied from our ABF-STEM images. The ABF-STEM images of tetragonal SRO ($β = 8$) (Figure 2e, Figures S4c and d) suggests that the local distortions of the $RuO_6$ octahedra, although significantly suppressed, are not exactly zero (> 2º). As $β$ increases within the tetragonal phases ($β > 8$), the small deviations could be further suppressed, leading to an enhancement in $T_c$.

With an increase in $β$, the uniaxial magnetocrystalline anisotropy ($K_u$) was observed to increase, whereas the saturation magnetization ($M_s$) was observed to decrease. Magnetic-field dependent magnetization ($M(H)$) measurements were performed at 5 K (Figure 3b) to characterize the FM hysteresis. The $M(H)$ curves of the SLs show small anomaly around zero $H$-field, which might originate from domain effect.[42,43] As is evident from the $M(H)$ curves, the coercive field ($H_c$) is related to $K_u$ along the $[001]_{pc}$ direction according to the relation of $H_c \leq 2K_u/\mu_o M_s$.[44, 45] Even though the $β$-dependent $M_s$ could also affect to $K_u$, the variation of $H_c$ was much larger than that of $M_s$ in the SLs. In general, the $K_u$ of SRO has been studied in the context of application to spintronics,[46] while most previous studies have focused on the modulation of lattice parameters for the engineering of $H_c$. In this study, we have demonstrated that $H_c$ varies significantly based on the extent of selective octahedral distortion and structural symmetry modification. As summarized in Figure 3d, the $H_c$ values were observed to increase with an increase in $β$, reaching ~1.7 T when $β = 18$ and 24. This value is more than eight times larger than that corresponding to single SRO thin films; further, it is comparable to that of rare-earth magnets used in high-density recording media.[47] On the other hand, the crystalline symmetry was also observed to determine the $M_s$ in SRO, as evidenced in Figure 3e. Herklotz and Dörr had theoretically predicted that the suppression of octahedral tilt quenches the magnetic moment of SRO, consistent with our observation.[21] In our SL systems, including the single SRO thin films, the orthorhombic phases were observed to exhibit the same $M_s$ value of ~1.7 $\mu_B$/Ru. In contrast, $M_s$ values of tetragonal SLs were observed to systematically decrease with an increasing STO thickness. SRO single film with modified crystalline symmetry consistently shows the suppressed $M_s$ in the tetragonal phase.[27] Yet, the experimental clues and physical interpretation of the microstructure-dependent $M_s$ is lacking. The decreasing trend of $M_s$ detected in the tetragonal symmetry is clearly contrary to the increasing trend of $T_c$, indicating that a simple FM model based on magnetic exchange interaction cannot explain the $\beta$-dependency of $M_s$ within tetragonal SRO.

As $\theta_{M\text{-}O\text{-}M}$ became flat, the local environment for the Ru orbital states was altered, providing a possible explanation for the structural dependence of $M_s$. Figure 4a shows a schematic diagram of the electronic structures and spin states corresponding to different structural symmetries of SRO. In general, orthorhombic SRO possesses a low spin state ($S = 1$), with four occupied $t_{2g}$ orbitals. On the other hand, tetragonal SRO with a flattened $\theta_{M\text{-}O\text{-}M}$ along $[001]_{pc}$ induces additional $t_{2g}$ splitting ($\Delta t_{2g} = E\,(d_{xy}) - E\,(d_{xz,yz})$) of the Ru 4$d$ orbitals, further altering the orbital occupation states. In particular, a larger $\Delta t_{2g}$ of the tetragonal SRO can partially change the $d_{xy}$ orbitals to the $d_{xz,yz}$ orbitals with an opposite spin, which would lead to a reduced $M_s$.[36] Although variations in the magnetic easy axis are also capable of influencing $M_s$, we confirmed that, in our case, the structural phase transition did not alter the magnetic easy axis, based on angle-dependent Hall measurements (data not shown). Figures 4b and S5 show the partial density of states (PDOS) of unoccupied $t_{2g}$ orbitals in the conduction band of SRO with orthorhombic and tetragonal structural symmetries. We fitted the DOS using the Lorentzian peak and obtained the center energy. Whereas the $d_{xy}$ and $d_{xz,yz}$ states in the orthorhombic symmetry were mostly degenerated (DOS ($t_{2g}^o$) in the top panel of Figure 4b), they showed a larger separation in the tetragonal symmetry with a clearly enhanced $\Delta t_{2g}$ (DOS ($t_{2g}^t$) in the central panel of Figure 4b). The difference between the respective PDOSs ($\sigma\,(t_{2g})$

= DOS ($t_{2g}^t$) − DOS ($t_{2g}^o$), bottom panel of Figure 4b) indicated an increase (decrease) in the number of unoccupied $d_{xy}$ ($d_{xz,yz}$) orbital states in tetragonal SRO, which is consistent with the scenario described in Figure 4a.

We employed X-ray absorption spectroscopy (XAS) to experimentally verify variations in the electronic structure effected via octahedral tilt penetration control (Figures 4c and S6). XAS revealed information regarding the excited electronic structures of Ru 4$d$ orbitals states, which is sensitive to the local atomic environment. The Ru $L_3$-edge XAS spectrum can be roughly attributed to the electron transitions from Ru 2$p$ core hole to Ru 4$d$ $t_{2g}$ (~2839.5 eV) and $e_g$ (~2842 eV) orbitals, respectively, although the final states are entangled owing to strong electron correlations (Figure S6a).[17] All the spectra consistently showed that the oxidation state of Ru was almost +4 with negligible energy shifts of $t_{2g}$ and $e_g$ manifolds. Additionally, Ti $L_3$-edge (458-462 eV) XAS spectra (Figure S6b) revealed the prevalence of only the Ti$^{4+}$ valence state, indicating no (unintended) external effects, such as charge transfer across the interface or defect formation within the SLs.[34] The $\theta_{M\text{-}O\text{-}M}$-dependent occupation in the Ru-$t_{2g}$ state led to the evolution of orbital polarization, which was observed via X-ray linear dichroism (XLD = $I_{x,y} - I_z$). Here, $I_{x,y}$ and $I_z$ denote the XAS intensities obtained via X-ray polarizations along the $x,y$- ([100]$_{pc}$ or [010]$_{pc}$) and $z$-axis ([001]$_{pc}$) directions, respectively (see the experimental section for further details). Each intensity reflects the electron excitation to the Ru 4$d$ orbitals in the direction of the X-ray polarization, as shown in the inset of Figure 4c. For instance, for the Ru-$t_{2g}$ states, $I_z$ reflects the transition to the $d_{xz,yz}$ orbitals, whereas $I_{x,y}$ reflects half the transitions to the $d_{xy}$ and the other half to the $d_{xz,yz}$ orbitals. Therefore, the sign of the XLD can be utilized to reveal the anisotropic orbital subshell state corresponding to each energy level. To assign the peak positions more rigorously, we simulated the XLD spectrum, as presented in the top panel (see the experimental section for further details) of Figure 4c, confirming the aforementioned argument. The experimental XLDs were shown in the bottom panel of Figure 4c. Whereas the XLDs of the $e_g$ states exhibit no significant changes in intensity, those of the $t_{2g}$ states clearly showed a systematic change depending on $β$. With an increase in $β$, the first peak at ~2839.5 eV became more intense, indicating an enhancement of unoccupied DOSs (i.e. a decrease of the population) of $d_{xy}$ orbital states. These results consistently support the crystalline symmetry-dependent spin states shown in Figure 4a, which would lower the total magnetization of SRO with tetragonal symmetry.

In conclusion, we controlled the propagation of octahedral tilt by atomically designing SRO/STO SLs. The selective manipulation of octahedral tilt in the SRO layer allowed us to study the effects of the crystalline symmetry on its electromagnetic properties, by isolating the influences of extrinsic origin such as strain relaxation, growth-induced defects or vacancies, or charge transfer across the hetero-interfaces. Furthermore, it provided us with another tuning knob of the functionality, enabling the electronic structure to be fine-tuned to modulate the desired ferromagnetic properties for future spintronic applications.

**Experimental Section**

*Thin film growth*: Atomically controlled [(SrRuO$_3$)$_\alpha$|(SrTiO$_3$)$_\beta$] superlattices ([$\alpha$|$\beta$] SLs) with $\alpha$ and $\beta$ number of atomic unit cells were synthesized using pulsed laser epitaxy on (001) STO substrates. Both SRO and STO layers were deposited at 750°C under 100 mTorr of oxygen partial pressure from the stoichiometric ceramic target using a KrF laser (248 nm; IPEX-868, Lightmachinery). We used a laser fluence of 1.5 Jcm$^{-2}$ and a repetition rate of 5 Hz. For the stoichiometric film growth, we used a high oxygen partial pressure, at which conventional reflection high energy electron diffraction cannot operate. Thus, we manipulated the number of u.c. of the SLs utilizing a customized automatic laser pulse control system programmed by LabVIEW. Based on the SL peaks in the XRD $\theta$-$2\theta$ scans, we characterized the thickness of the SL period using Bragg's law, as follows:

$$\Lambda = \frac{n\lambda}{2}(sin\ \theta_n - sin\ \theta_{n-1})^{-1}, \tag{1}$$

where $\Lambda$, $n$, $\lambda$, and $\theta_n$ denote the period thickness, SLs peaks order, wavelength of the X-ray, and $n$ th-order SL peak position, respectively. All of the layers showed a small thickness deviation of <1 u.c. (≈0.4 nm). Atomic-scale STEM images consistently supported our thickness control techniques.

*Lattice structure characterization:* High-resolution X-ray diffraction (HRXRD) measurements were performed using a Rigaku Smartlab and a PANalytical X'Pert X-ray diffractometer. Atomic-scale imaging of SLs was performed on a spherical aberration-corrected scanning transmission electron microscope (STEM; ARM200CF, JEOL) operating at 200 kV. To detect the $\beta$-dependency of octahedral distortions in SRO layers, the annular bright-field (ABF) imaging mode was employed along with the high-angle annular dark-field (HAADF) imaging mode. The incident electron probe angle was set to 23 mrad, giving rise to a probe size of 0.78 Å. The ABF and HAADF signals were simultaneously collected over detector angle ranges of 7.5 – 17 and 70 – 175 mrad, respectively. Cross-sectional thin samples for STEM analysis were prepared using a dual-beam focused ion beam system (FIB, FEI Helios Nano Lab 450); subsequently, low-energy Ar ion milling at 700 V (Fischione Model 1040, Nanomill) was carried out for 15 min to remove surface layers damaged owing to heavy Ga ion beam milling in the FIB system.

*Magnetization measurement*: Temperature-($M$ ($T$)) and magnetic field-dependent magnetization ($M$ ($H$)) were measured using a Magnetic Property Measurement System (MPMS, Quantum Design). The measurements were performed at a range of 300 to 2 K under 100 Oe of the magnetic field along the out-of-plane direction of the thin films. $M$ ($H$) curves were obtained at 5 K with a magnetic field along the out-of-plane direction.

*XAS measurement*: Ru $L_3$-edge XAS was performed at the 16A1 beamline of the Taiwan Light Source in the fluorescence yield mode at room temperature, whereas Ti $L_{2,3}$-edge XAS was performed in the 2A beamline of the Pohang Light Source in the total electron yield mode at room temperature. The probing depth of Ru $L$-edge XAS was approximately a micron, far exceeding the total thickness of the SLs, whereas that corresponding to Ti $L$-edge XAS was in the order of 10 nm. To obtain the polarization-dependent data, the samples were either set in a beam-normal geometry ($I_{x,y}$) or rotated by 70° [($\cos^2 70° \times I_{x,y}$) + ($\sin^2 70° \times I_z$)].

*XLD simulation and peak assignment*: To enable a clear peak assignment, we simulated the XLD spectrum for a hypothetical orthorhombic SRO model using a charge transfer multiplets calculation code, CTM4XAS.[48] In the model, the atomic multiplets of $d^4$ many-body states under crystal fields of $D_{4h}$ point symmetry were considered in the scheme of configuration interactions with charge transferred states. All the values of the parameters (for instance, crystal field splitting energies, 10Dq, Ds and Dt, the transfer matrix, and the charge transfer energy) were adopted from reference,[49] except for the values of the Slater integrals, which were reduced to ~50% of the atomic values to account for the itinerant nature of the $d$ electrons in SRO. In the ground state ($d^4$; $S = 1$), the first unoccupied orbital state was $d_{xz,yz}$. Thus, the lowest energy feature in the XLD spectrum for orthorhombic SRO should appear as a dip for $d_{xz,yz}$. A peak for $d_{xy}$, a dip for $d_{z2}$, and a peak for $d_{x2-y2}$ should follow in the order of increasing energy. Meanwhile, in the case of tetragonal SRO, the first dip for $d_{xz,yz}$ (~2838.5 eV) apparently disappeared, and the peak for $d_{xy}$ (~2839.5 eV) increased in intensity because of the slight increase (decrease) in the number of electrons at the $d_{xz,yz}$ ($d_{xy}$) orbital, which is consistent with the scheme shown in Figure 4a.

*DFT calculations*: Our first-principles DFT calculations were performed using generalized gradient approximation (GGA)[50] and the projector-augmented wave method with a plane-wave basis,[51] as implemented in the Vienna ab-initio simulation package (VASP) code.[52] For the Brillouin-zone integration, we used a kinetic energy cutoff of 500 eV and Γ-centered 8 × 8 × 8 $k$-point meshes. For the DOS calculations, we considered orthorhombic *Pbnm* (Glazer notation, $a^-a^-c^+$), and tetragonal *P4/mmm* ($a^0a^0c^0$) structures composed of 20 atoms, and their in-plane lattice parameter was fixed to be $\sqrt{2}a_{STO}$. To consider on-site Coulomb interactions, a Hubbard $U$ of 1.6 eV was applied to the Ru-$d$ orbital for all calculations.[53] The calculations were converged in energy to $10^{-6}$ eV cell$^{-1}$, and the structures were allowed to fully relax until the forces reduced below $10^{-3}$ eV Å$^{-1}$.

**Acknowledgements**
This work was supported by the Basic Science Research Programs through the National Research Foundation of Korea (NRF) (NRF-2019R1A2B5B02004546 and NRF-2018R1D1A1B07043427). J.L. and Y.-M.K were supported by the Samsung Research Funding & Incubation Center of Samsung Electronics (SRFC-MA1702-01). Y.-M.K. was supported by the IBS (IBS-R011-D1).

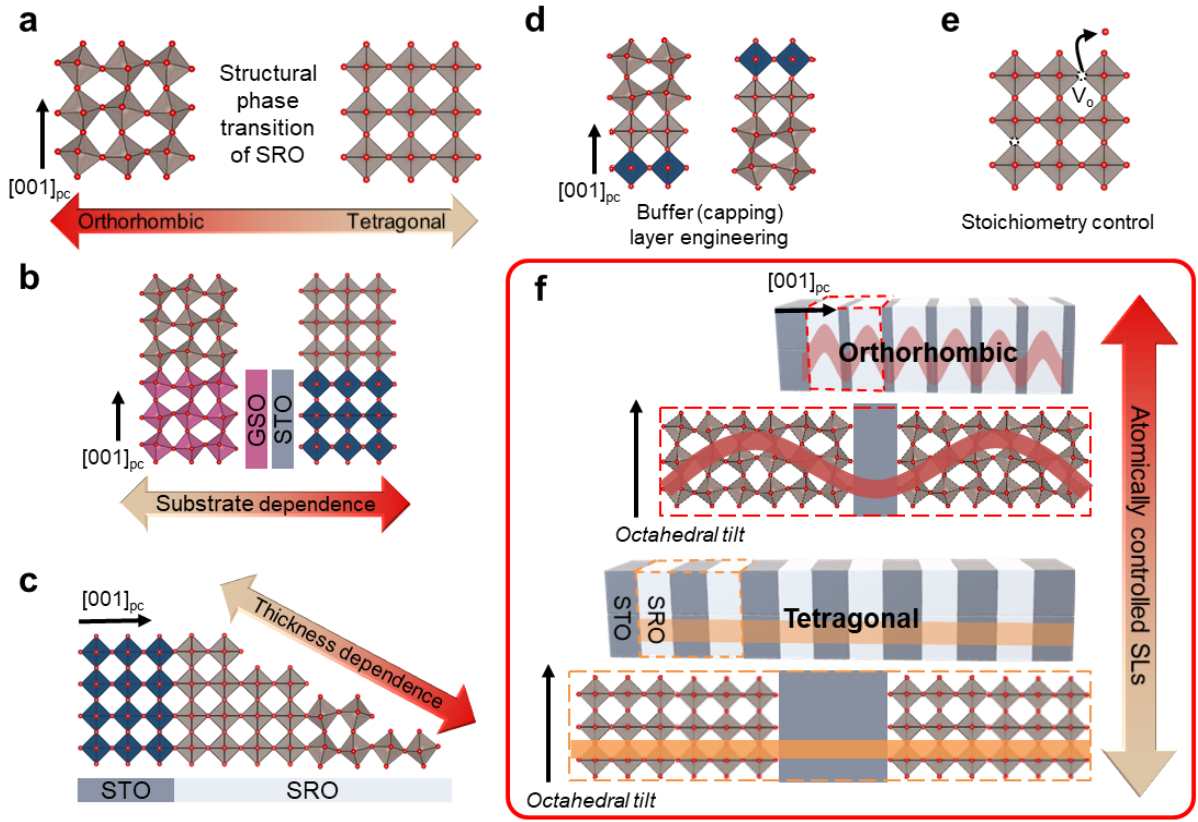

**Figure 1.** Customization of the tilt of $RuO_6$ octahedra in SRO crystals. a) Schematic representation of the structural phase transition of SRO from orthorhombic to tetragonal. Customization of the tilt of $RuO_6$ octahedra has been demonstrated by b) substrate epitaxial strain dependence,[23, 24] c) thickness dependence,[23] d) buffer (capping) layer engineering,[26, 27] and e) stoichiometry control.[16] $V_o$ indicates oxygen vacancy. f) Sketch of octahedral tilt penetration control via atomically controlled SLs. The amplitudes of the red and orange lines indicate the extent of octahedral tilt. Note that $[hkl]_{pc}$ denotes the crystallographic orientations within the conventional perovskite pseudo-cubic notation.

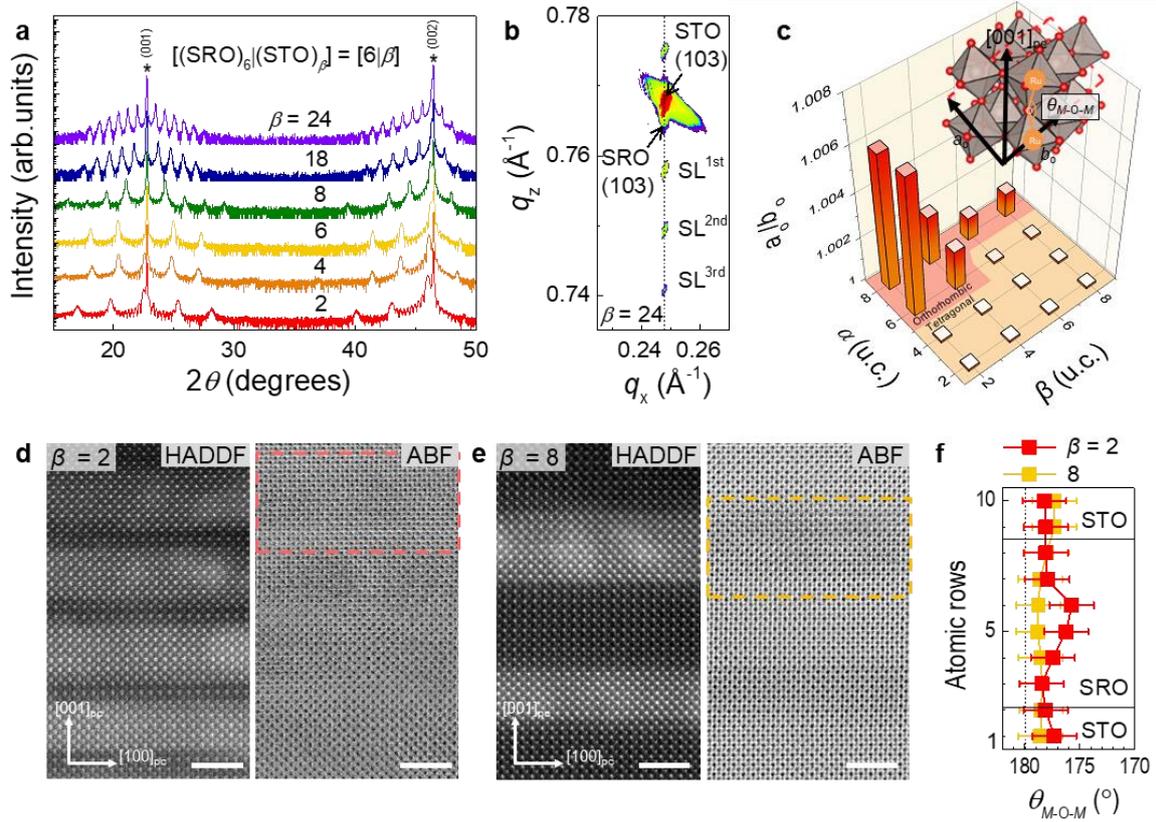

**Figure 2.** Structural phase transition via modulation of RuO$_6$ octahedra in atomically controlled SRO/STO SLs. a) X-ray diffraction $\theta$-$2\theta$ scans were shown for the well-defined [(SRO)$_6$|(STO)$_\beta$]$_{10}$ SLs grown on STO substrates. The asterisk (*) indicates the STO substrate peaks. b) Reciprocal space map of the SL with $\beta = 24$, around the (103) Bragg reflection of the STO substrate, indicating the fully strained SL (SL$^{\pm N \text{th}}$) with a coherent in-plane lattice parameter as that of the substrates. c) The structural phase map as functions of $\alpha$ and $\beta$. The inset schematically depicts orthorhombic distortions ($a_o/b_o$) extracted from the lattice parameters of the orthorhombic unit cell. The $\alpha$ and $\beta$ indicate the atomic u.c. of SRO and STO layers. The $\theta_{M\text{-}O\text{-}M}$ is $M$-O-$M$ bonding angle. The red (orange) region indicates the orthorhombic (tetragonal) phase. The HAADF- (left) and ABF-STEM (right) results are shown for the SLs with d) $\beta = 2$ and e) 8. The scale bars denote 2 nm. f) $\theta_{M\text{-}O\text{-}M}$ of SLs are extracted by averaging 34 oxygen displacements along the in-plane direction of the dotted rectangles of STEM images. Note that it was not possible to obtain the octahedral rotation along the in-plane direction based on the current experimental configuration.

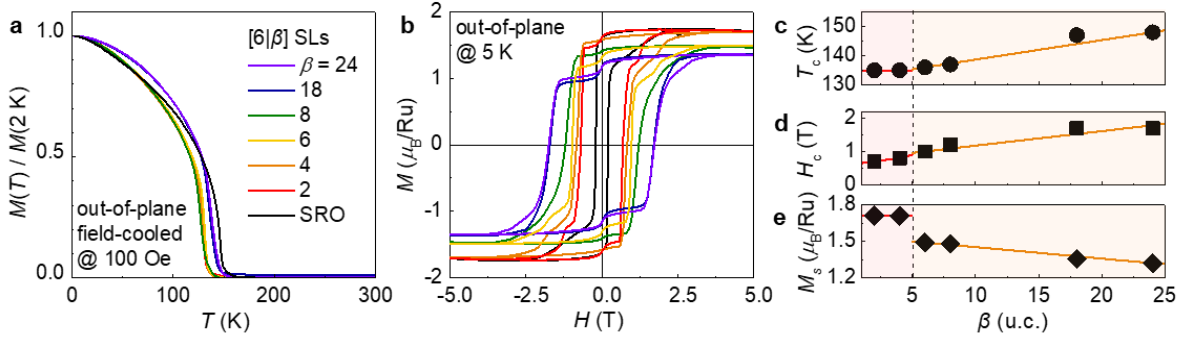

**Figure 3.** Ferromagnetic properties of SRO/STO SLs tuned by the octahedral tilt propagation. a) Field-cooled $M(T)$ of [6|$\beta$] SLs with different $\beta$ have been characterized at 100 Oe, along the out-of-plane direction. b) $M(H)$ curves of the SLs were obtained at 5 K. c) $\beta$-dependent FM transition temperature ($T_c$) of SLs is extracted from $M(T)$. d) Coercive field ($H_c$) and e) saturation magnetization ($M_s$) at 5 T, extracted from $M(H)$ curves as functions of $\beta$. The vertical dashed line represents the border of the structural phase transition of the SLs depending on STO thickness. The red (orange) region indicates the orthorhombic (tetragonal) structure of SLs. The solid line is a linear fit of the data points within each crystalline symmetry as a guide to the eye.

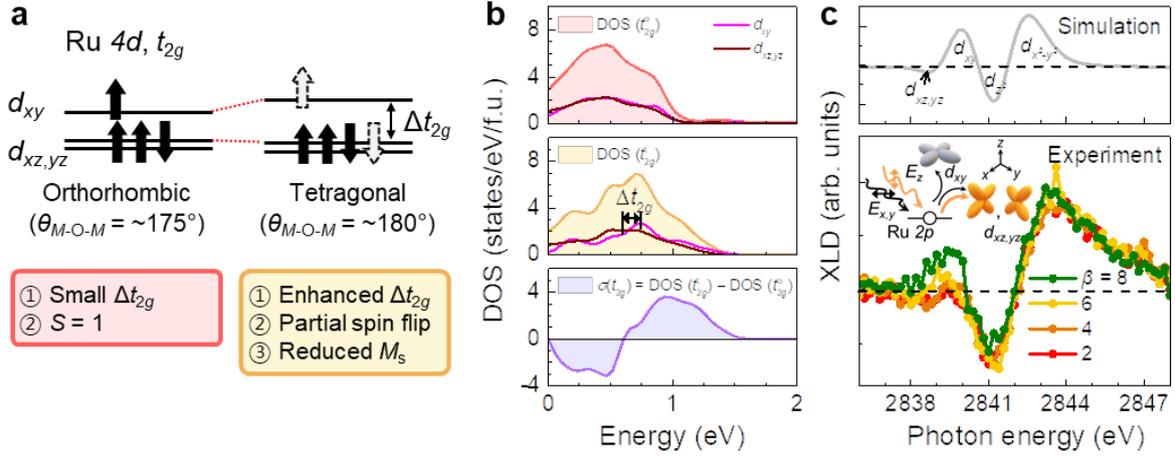

**Figure 4.** Structure-dependent electronic structure of SRO/STO SLs. a) Schematic diagram of possible Ru-$t_{2g}$ orbital states with different structural symmetries. Black solid (dashed white) arrows represent fully (partially) occupied spin states. The structure-dependent properties have been summarized in the panels below. b) Orbital selective partial density of states (PDOS) in the conduction bands of Ru-$t_{2g}$ states for orthorhombic (DOS ($t_{2g}^o$) in the top panel) and tetragonal SRO (DOS ($t_{2g}^t$) in the middle panel). Degenerated $t_{2g}$ states of orthorhombic symmetry can split into the $d_{xy}$ and $d_{xz,yz}$ states with an enhanced $\Delta t_{2g}$ in tetragonal symmetry. The bottom panel shows the difference of DOS ($\sigma(t_{2g})$) between tetragonal and orthorhombic symmetry. c) Linear dichroisms, $[I_{x,y} - I_z]$, have been gauged via simulated (top panel) and experimental (bottom panel) results. The inset shows the schematic of the polarization dependence. $E_{x,y}$ and $E_z$ denote the $x,y$- ($[100]_{pc}$ or $[010]_{pc}$) and $z$-directional ($[001]_{pc}$) electric polarization of the incident beam, respectively.

# Supporting Information

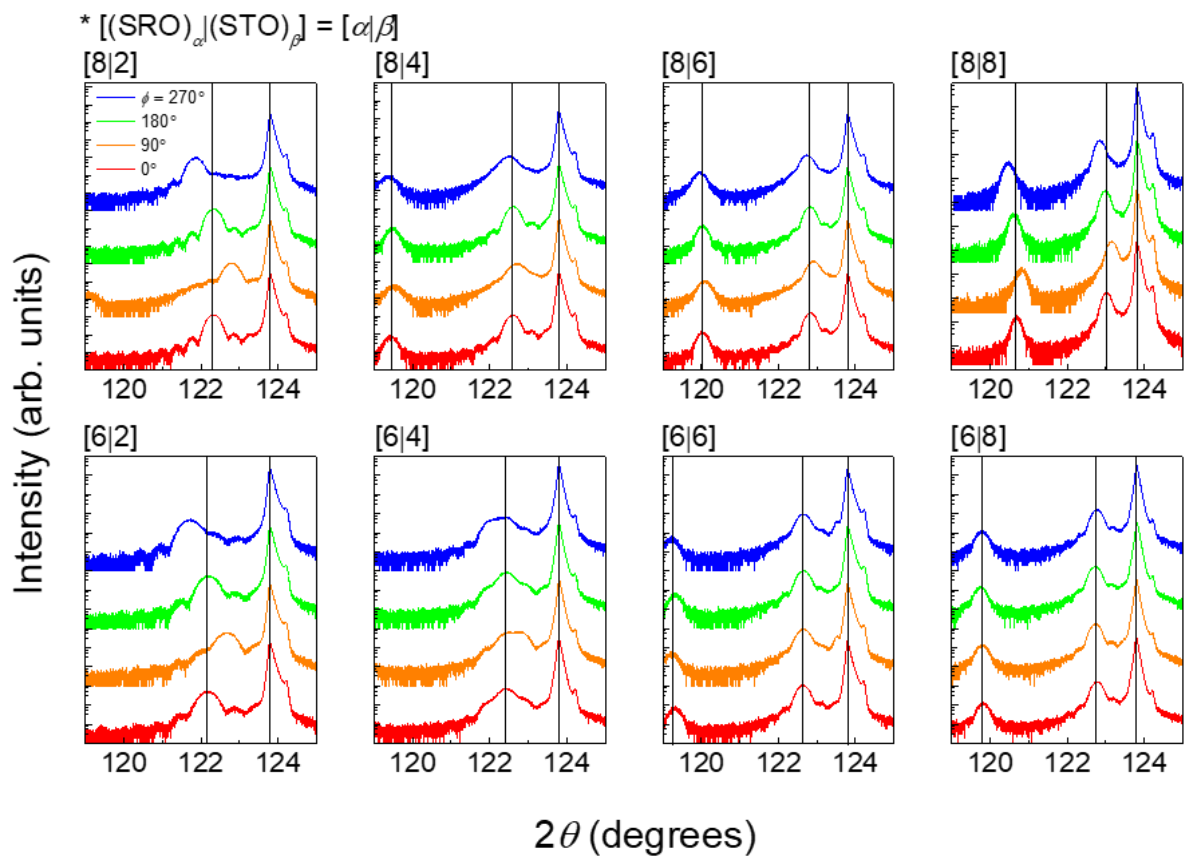

**Figure S1.** Structural characterization of the [α|β] SLs. Off-axis X-ray diffraction measurements for the [α|β] SLs around the STO (204) Bragg reflections with $\varphi$ angles of 0, 90, 180, and 270°. The vertical lines are guides to the eye.

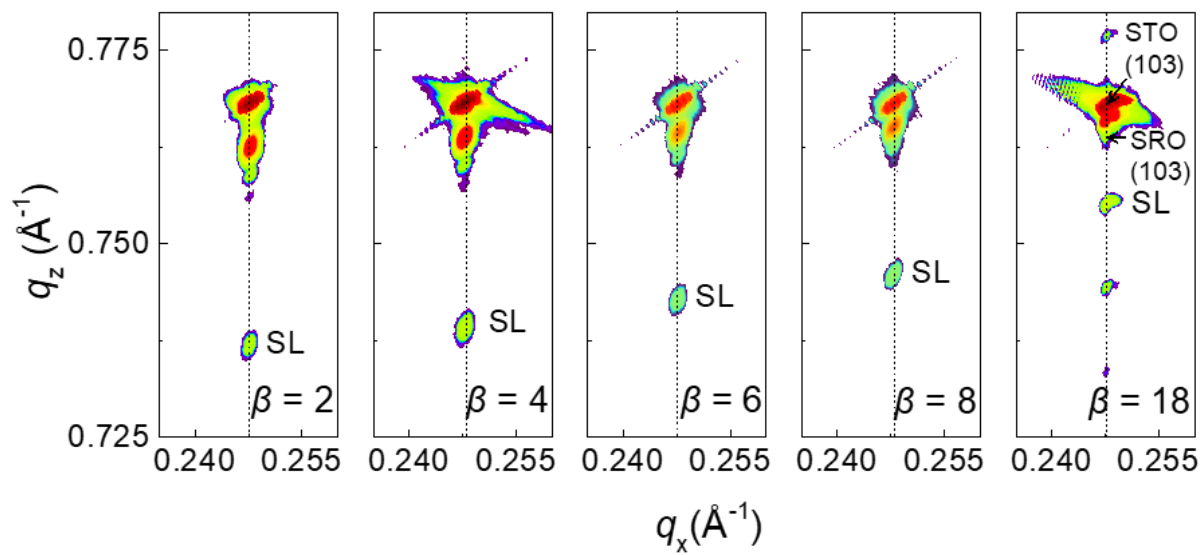

**Figure S2.** Epitaxial strain of [6|β] SLs. XRD RSMs of the SLs, shown for the [6|β] SL around the (103) Bragg reflection of the STO substrate, indicating the fully strained state of the SLs with the coherent in-plane lattice constant as that of the substrates.

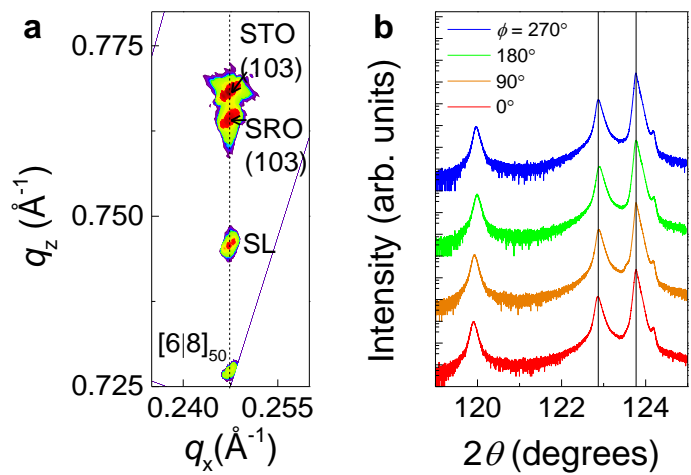

**Figure S3.** XRD results of [6|8]$_{50}$ SL, 6 u.c. layers of SRO and 8 u.c. layers of STO repeated 50 times along the growth direction, are shown. a) RSM and b) off-axis measurements consistently indicate that the tetragonal SRO is well maintained up to ~120 nm of SRO thickness with a fully strained state.

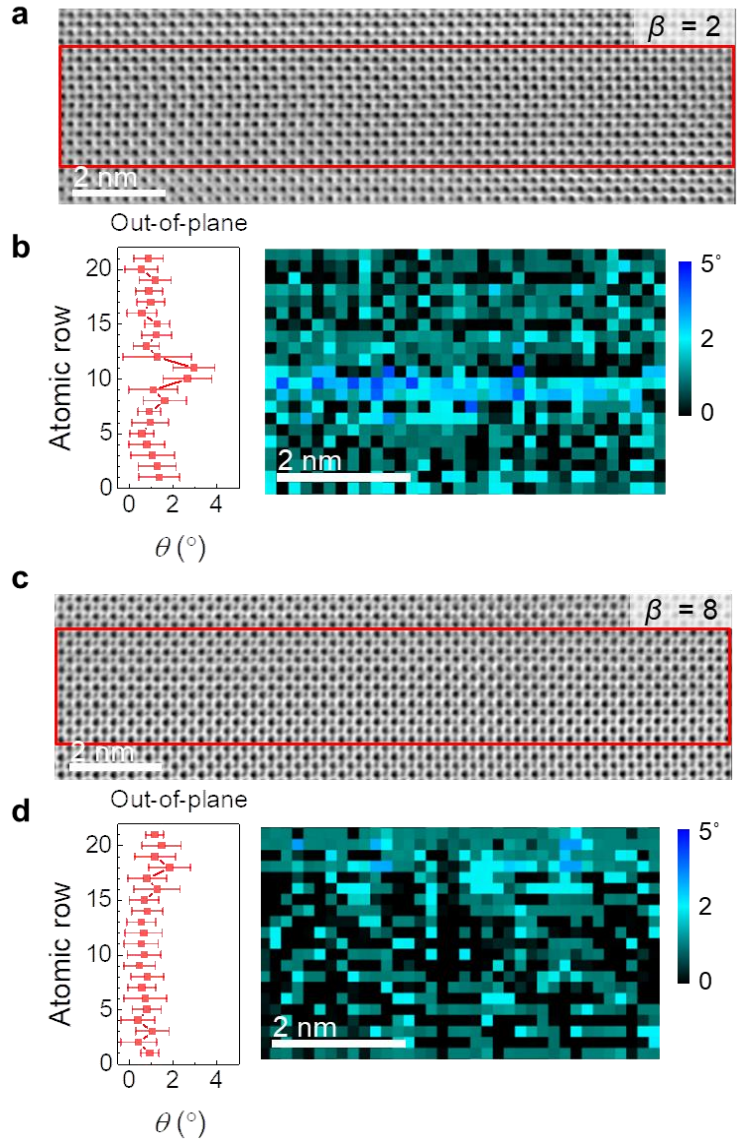

**Figure S4.** ABF-STEM images of [6|$\beta$] SLs in high magnification. ABF-STEM observation also displays the well-defined epitaxy of the SLs with a) $\beta$ = 2, and c) 8, with a clear visualization of the oxygen atoms, respectively. We extracted the average *M*-O bonding angles ($\theta$) (left panel) of the SLs with b) $\beta$ = 2, and d) 8, along the out-of-plane direction from the contour plot (right panel), respectively.

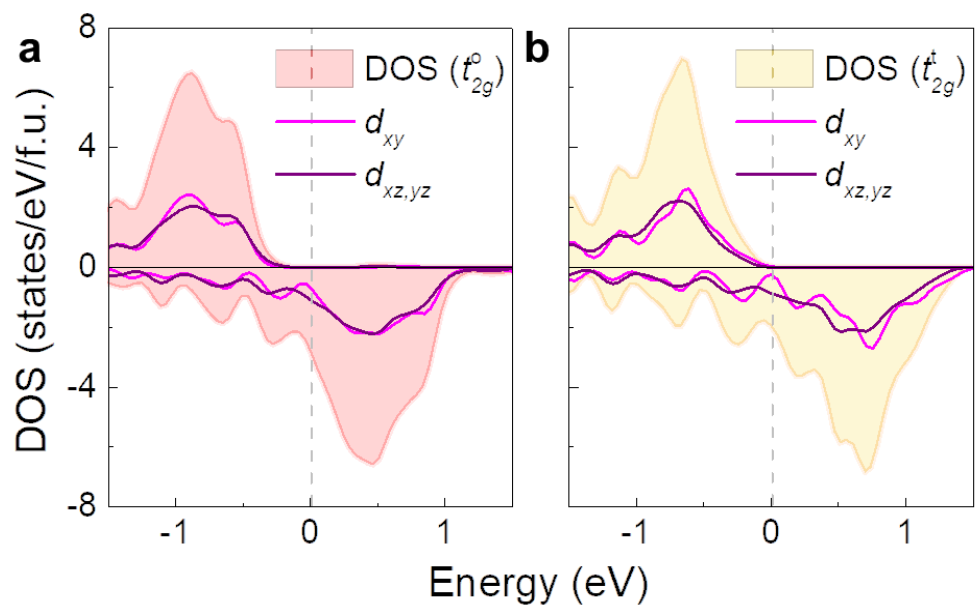

**Figure S5.** Orbital selective PDOS of Ru-$t_{2g}$ states for a) orthorhombic (DOS ($t_{2g}^o$)) and b) tetragonal SRO (DOS ($t_{2g}^t$)). The vertical dashed lines indicate the Fermi level.

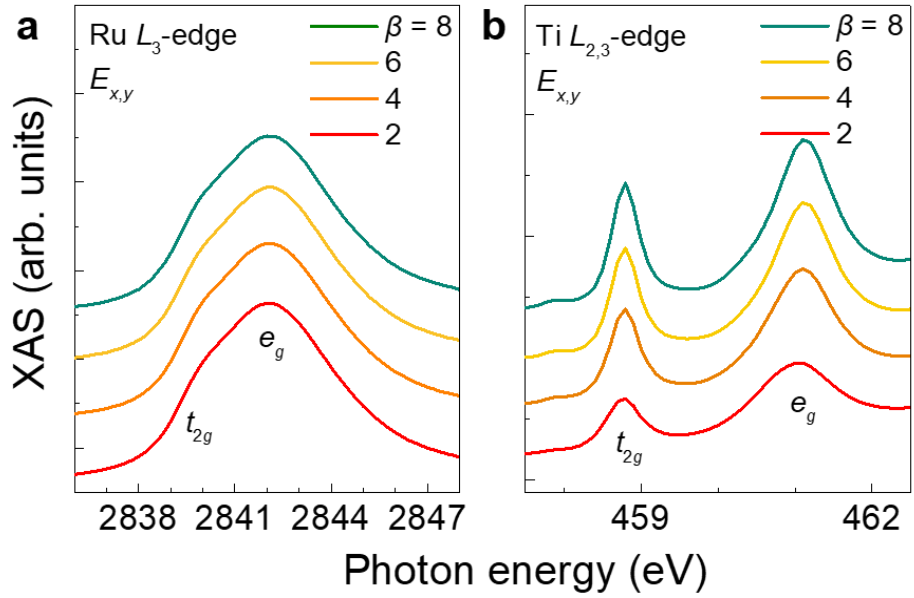

**Figure S6**. XAS spectra for the [6|$\beta$] SLs. a) Ru $L_3$-edge XAS spectra of [6|$\beta$] SLs with different $\beta$ values are acquired using a normal incident beam into the film surface ($E_{x,y}$), in which the electrical field of X-rays lies along the vertical direction in the measurement chamber. The $t_{2g}$ and $e_g$ energy levels of the Ru $L_3$-edge are assigned at ~2839.5 eV and ~2842 eV, respectively. b) The $t_{2g}$ and $e_g$ energy levels of the Ti $L_{2,3}$-edge are assigned at ~458.8 eV and ~461.1 eV, respectively.